\documentclass[%
 reprint,
 amsmath,amssymb,
 aps,prl,floatfix,showpacs
]{revtex4-1}

\usepackage{graphicx}
\usepackage{dcolumn}
\usepackage{epstopdf}
\usepackage{color}
\usepackage{bm}

\begin{document}

\newcommand{\LNO}{$\rm{LiNbO_3}$\,}
\newcommand{\PT}{$\rm{PbTiO_3}$\,}
\newcommand{\etal}{{\em et al.}}{}
\newcommand{\fig}[1]{Fig.~\ref{#1}}
\newcommand{\tab}[1]{{Table ~\ref{#1}}}
\newcommand{\AN}{$\rm{\AA}$\,}{}
\newcommand{\UPS}{$\rm{C/m^2}$\,\,}
\newcommand{\UPSS}{$\rm{\mu C/cm^2}$\,\,}
\newcommand{\UPY}{$\rm{C/m^2K}$\,\,}
\newcommand{\UPYY}{$\rm{\mu C/m^2K}$\,\,}
\newcommand{\ww}{0.47}
\newcommand{\Tc}{$\rm{T_c}$\,}
\newcommand{\Ts}{$\rm{T_s}$\,}
\newcommand{\Tt}{$\rm{T_1}$\,}
\newcommand{\Ths}{$\rm{T_{hs}}$\,}
\newcommand{\Ttri}{$\rm{T_{tri}}$\,}
\newcommand{\Tstar}{$\rm{T^*}$\,}
\newcommand{\degree}{\ensuremath{^\circ}}

\title{Giant electrocaloric effect around \Tc}

\author{Maimon C. Rose}
\email{mrose@gl.ciw.edu}
\author{R.~E. Cohen}%
\email{cohen@gl.ciw.edu}
\affiliation{%
Geophysical Laboratory, Carnegie Institution of Washington, 5251 Broad Branch Rd., N.W., Washington,  D.C. 20015}%

\date{\today}

\begin{abstract}
We use molecular dynamics with a first-principles-based shell model potential to study the electrocaloric effect (ECE) in lithium niobate, \LNO, and find a giant electrocaloric effect along a line passing through the ferroelectric transition. With applied electric field, a line of maximum ECE passes through the zero field ferroelectric transition, continuing along a Widom line at high temperatures with increasing field, and along the instability that leads to homogeneous ferroelectric switching below \Tc with an applied field antiparallel to the spontaneous polarization. This line is defined as the minimum in the inverse capacitance under applied electric field. We investigate the effects of pressure, temperature and applied electric field on the ECE. The behavior we observe in \LNO should generally apply to ferroelectrics; we therefore suggest that the operating temperature for refrigeration and energy scavenging applications should be above the ferroelectric transition region to obtain large electrocaloric response. The relationship among \Tc, the Widom line and homogeneous switching should be universal among ferroelectrics, relaxors, multiferroics, and the same behavior should be found under applied magnetic fields in ferromagnets.
\end{abstract}

\maketitle
The electrocaloric effect (ECE), the reversible temperature change of a polarizable material under the application and/or removal of an electric field, has the potential for efficient solid-state refrigeration\cite{bai2010}, but fundamental questions about the nature of the ECE have plagued the field for at least 50 years \cite{jona1962,scott2011}.  We  show how the ECE can be large well above \Tc, in spite of the apparent thermodynamic argument that it vanishes above \Tc . The ECE is the change in entropy with electric field, and thus can be used to pump heat or scavenge energy. Experiments show temperature changes of up to 12 K on application of reasonable electric fields of 480 kV/cm in PZT \cite{mischenko2006a,mischenko2006b}. Simulations with an effective Hamiltonian of the ECE show good agreement with experiments,\cite{prosandeev2008,lisenkov2009}. We study the behavior of \LNO as a simple, uniaxial ferroelectric \cite{Glass1976,Lines1977} for which we have a well-tested first-principles based interatomic potential\cite{sepliarsky2002,sepliarsky2004,sepliarsky2011,peng2011}, with which we previously studied the converse effect, pyroelectricity, under zero applied field\cite{peng2011}. Recently the ECE in Ba$_{0.5}$Sr$_{0.5}$TiO$_3$ was simulated using an effective Hamiltonian at constant volume.\cite{Ponomareva2012} Here we use an interatomic potential which includes all degrees of freedom to perform simulations at constant pressure and temperature, and develop a different picture of the ECE from Ref.~\onlinecite{Ponomareva2012}. We carefully study the ECE around \Tc, and find that ferroelectric switching below \Tc is closely related to the polarization response above \Tc under applied fields.

\begin{figure}[htbp]
\centering
\includegraphics[width=2.9in,angle=-90]{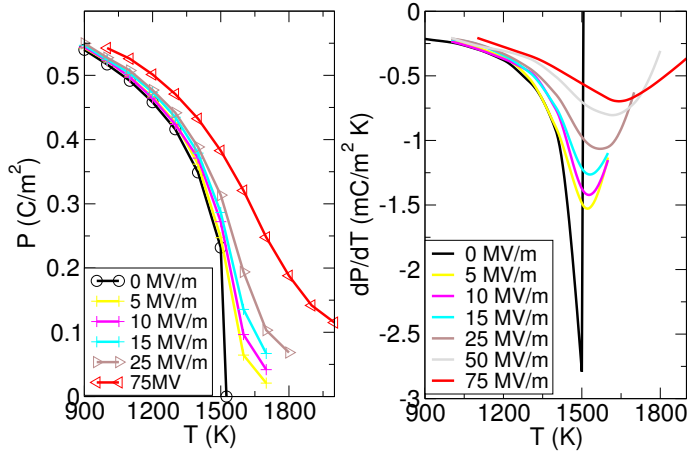}
\caption{{\label{dPdT}}(a) Average polarization of \LNO with applied field along the same direction as the spontaneous polarization and (b) derivative of polarization with respect to temperature. The Widom line \Tstar above \Tc is traced out by the inflections in $P$ or minima in $dP/dT$.}
\end{figure}

We performed molecular dynamics (MD) simulations on supercells of 8x8x8 unit cells, with a total of 5210 atoms using the DLPOLY package \cite{dlpoly} with a Nos\'{e}-Hoover thermostat and barostat in the N$\sigma$T ensemble. The shell model potential was fit to first-principles density functional theory (LDA) energies, forces, stress, effective charges, dielectric constants, phonons frequencies and eigenvectors and has proven predictive and robust.  At lower temperatures, long runs were needed  to reach equilibrium; e.g.~at 800 K, 22.5 ns was required to equilibrate the shell and atomic positions. Once equilibrium was reached, the stable configuration was used as a starting point for runs at similar temperatures. For all simulations with the electric field applied in the direction opposite to the polarization, the simulations were either 44 or 110 ps long with a timestep of 0.2 and 0.5 fs, respectively. At 4 or 10 ps intervals the electric field was increased by 5 MV/m. For simulations with the field applied parallel to the spontaneous polarization, we ran a simulation at a fixed applied field and temperature for 100-200 ps with a timestep of 0.5 fs. All averages were taken after the equilibration period, which was typically 25-50 ps. Simulations with applied pressure ran for 100 ps with a timestep of 0.5 fs. We employ the convention defining fields applied in the same direction as the spontaneous polarization as {\it positive} or {\it parallel} and the fields applied in the direction opposite that of the spontaneous polarization as {\it negative} or {\it antiparallel}. The results of polarization versus temperature and applied field are shown in Fig.~\ref{dPdT}. We find the zero-field ferroelectric to paraelectric transition to be 1506K, which slightly overestimates the literature value of ~1480K\cite{Glass1976}. 

\begin{figure}[htbp]
\centering
\includegraphics[width=3.5in]{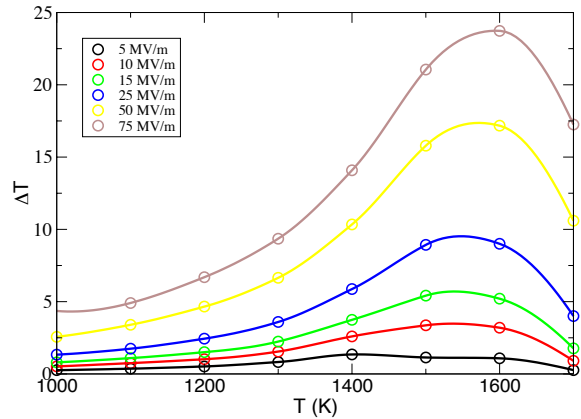}
\caption{\label{deltaT}Calculated change in temperature as a function of temperature for various applied electric fields.}
\end{figure}

To investigate the electrocaloric effect we used the Maxwell relation
\begin{equation}{\label{maxwell1}} 
\left(\frac{\partial S}{\partial E}\right)_T=\left(\frac{\partial P}{\partial T}\right)_E
\end{equation}
or
\begin{equation}{\label{maxwell2}}
\left(\frac{\partial T}{\partial E}\right)_S=\frac{T V}{C_{p,E}}\left(\frac{\partial P}{\partial T}\right)_E  ,
\end{equation}
where $P$ is the macroscopic polarization, $E$ is the macroscopic electric field, $C$ is the heat capacity, $V$ is the volume, $T$ is the absolute temperature, and $p$ is the pressure, which when integrated gives 
\begin{equation}{\label{maxwell3}}
\Delta T = - \int_{0}^{E} \frac{T V}{C_{p,E}}\left(\frac{\partial P}{\partial T}\right)_E dE .
\end{equation}
To calculate the change in the temperature, equation \ref{maxwell2} can be solved numerically using Mathematica.  For our case approximating the solution using eq.~\ref{maxwell3} by fixing $T$ on the right hand side gives an error of less than 1\% in $\Delta T$, which is less than errors from the model potential, LDA, etc. The maximum $\Delta T$ is found at progressively higher temperatures as the strength of the field is increased (Fig.~\ref{deltaT}). We also performed constant energy (NVE) simulations with applied fields, and found agreement within error to the results in Fig.~\ref{deltaT}, obtained from Eq.~\ref{maxwell3}, considering the differences between constant V and constant p simulations. 

For small negative or antiparallel fields below the ferroelectric transition we expect to find the same values of the ECE as for small positive fields since $(\frac{\partial P}{\partial T})|_E$ is continuous around $E=0$. When we apply a negative field larger than the coercive field $E_c$ at a given temperature \Ths, the polarization switches homogeneously, and the direction of the polarization of the crystal becomes parallel with the direction of the applied field (see page 79-80 in Ref.~\onlinecite{landau} for a discussion of this instability). We call this temperature for a given electric field the homogenous switching temperature, \Ths. Normally switching occurs through growth of preexisting domains with a more preferred polarization direction, or nucleation and growth of new domains. Homogeneous switching, with no preexisting domains, is the limit of metastability of the energetically unfavored polarization direction in an applied field. We recently studied homogeneous switching in detail in \PT \cite{zeng2011}. This switching occurs, depending on the strength of the applied negative field, on the time scale of our simulations between 4 and 10 ps.

We performed constant energy (NVE) simulations to study the temperature change during polarization switching. As expected, the temperature goes up sharply with switching: since the potential energy drops, the kinetic energy, and thus T, increases. By varying the applied field frequency, however, it should be possible to get the polarization switching to go 180\degree's out of phase and make a refrigeration cycle based on polarization switching. Such behavior may be an explanation of the dependence of ECE on applied field frequency seen in experiments \cite{bai2010}.

In order to understand the line of maximum ECE, we consider the thermodynamics of ferroelectrics in an applied electric field. At zero field the ferroelectric transition, \Tt, is first-order, and with applied field, \Tt increases as $\frac{dT_1}{dE}=-\frac{\Delta P}{\Delta S}$ \cite{akcay2007}. With increasing applied field, a tricritical point, \Ttri, is reached where the transition becomes second-order, \Ts. In BaTiO$_3$ this first-order region between \Tt and \Ttri is 18K.\cite{scott2011} With a further increase in field a critical point \Tc is reached, and there is no phase transition above that temperature and applied field, E$_c$.  We were not able to resolve the region between \Tt and \Tc in our MD simulations at zero pressure.  Although there is no phase transition above \Tc for higher fields, there is still an inflection point in $P$ versus $T$ and a line is traced out with increasing $E$ that can be considered the Widom line for this problem \cite{xu2005,simeoni2010,kutnjak2006,kutnjak2007}. We label the Widom line \Tstar. At  temperatures below \Tt, negative fields greater in magnitude than the coercive field switch the polarization, tracing the line of homogeneous switching, \Ths. This instability is found clearly in first-principles computations at constant electric field and zero temperature in \PT (see Fig. 6c in Ref.~\onlinecite{Hong2011}). The importance of the inverse capacitance $C^{-1} \propto \epsilon^{-1}$, which is the second derivative of the internal energy with electric displacement field D, where $\epsilon$ is the dielectric constant as discussed in Ref. \onlinecite{Stengel2009}. We propose a line of minima in the inverse capacitance $C^{-1}$ joins \Ths, \Tt, \Ttri, \Tc, and \Tstar. Below \Tc, $C^{-1}=0$ along this line, whereas it is a minimum above \Tc defining \Tstar. These temperatures are sketched in the inset to Fig.~\ref{Tc3} below. 

As J.F. Scott discusses, the issue of ECE above \Tc has been a argument spanning many decades over whether the ECE is expected above or below \Tt \cite{scott2011}. The confusion stems from misconstruing the Maxwell relation (Eq. 1). Above \Tt, $P=0$ at zero field, implying zero ECE, but at larger fields $P \neq 0$ so that there is an ECE in any material, not just a ferroelectric. Furthermore there is no requirement of an underlying ferroelectric phase transition. The fact that the dielectric constant $ \epsilon \propto C$ is high above \Tt in a ferroelectric, however, makes the application of ferroelectrics above \Tt optimal, rather than far below \Tt as seems to be standard practice now. The susceptibility is maximal along the Widom line, so that is where we expect to see the greatest ECE, as we find. The maximum in $\Delta T$ occurs along the line defined by $\min C^{-1}$, and our results clarify the issue of ECE above \Tt. The order parameter $Q$ for polarization correlations $Q=\lim_{r\rightarrow \infty}<P(r)P(0)>|_E$, and the maxima in the ECE should be where $\frac{d^2Q}{dE dT}$ is maximal. 

Ponomareva and Lisenkov state that ``$\Delta T$ peaks near the transition temperatures which is the consequence of the larger configurational disorder associated with these points" \cite{Ponomareva2012}. To the contrary, we find that $\Delta T$ continues to increase as the electric field increases, well above the zero field \Tt. So we find that operation above \Tt would be advantageous in ECE applications. The issue of disorder will be discussed further below.

\begin{figure}[htbp]
\centering
\includegraphics[width=3.5in]{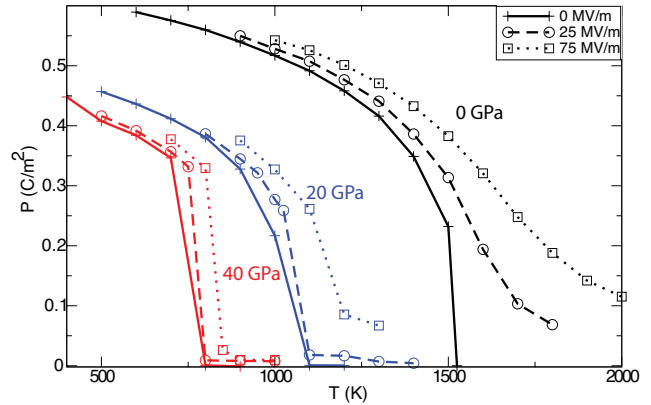}
\caption{\label{pressure}Effect of pressure and electric field on polarization in \LNO . Note the strong decrease in the transition temperatures with pressure, and the increase in the phase transition region which appears to be first order up to at least 25 MV/m and about 100K at 20 GPa. At zero pressure the phase transition disappears at relatively small fields.}
\end{figure}

We have also simulated \LNO as a function of applied pressure.  As pressure increases, \Tt strongly decreases, so that it is 750K at 40 GPa.  As pressure increases and \Tt decreases, we find consistency with the picture above, with the ECE maximal along the Widom line.  Unexpectedly, we found that pressure increases the phase transition region between \Tt and \Ttri under applied field, so that it is more easily discernible (Fig.~\ref{pressure}).  We also find a structural phase transition between 60 and 70 GPa, consistent with the transition  seen experimentally at about 40 GPa\cite{jayaraman1986,suchocki2006}.

\begin{figure}[htbp]
\centering
\includegraphics[width=3.5in]{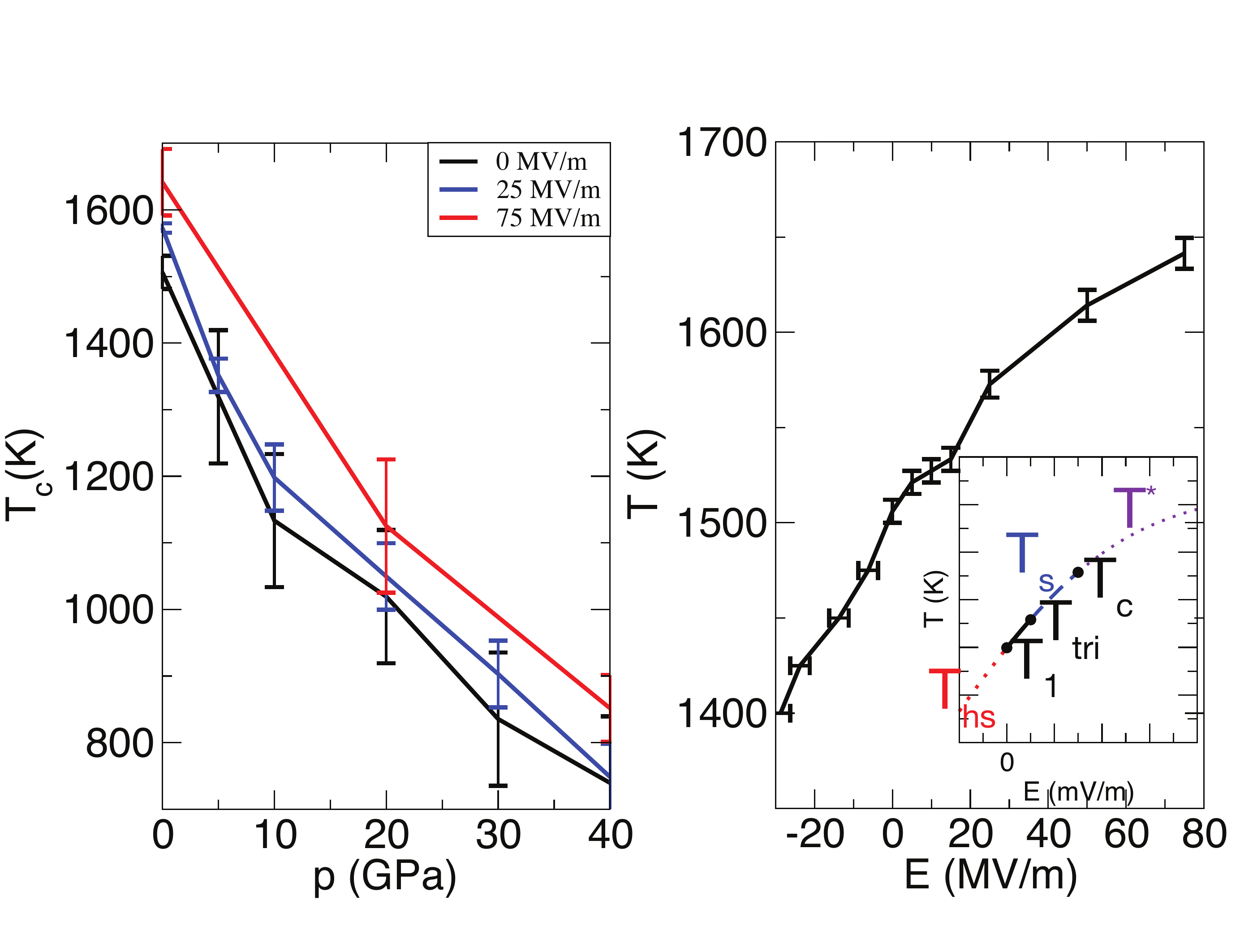}
\caption{\label{Tc3}Effects of pressure and electric field on the transitions in \LNO. \Tt, \Tc, \Tstar along the Widom line, and \Ths, the homogeneous switching temperature fall along a smooth curve. The right hand figure shows the line of minimum inverse capacitance, and the inset shows the relationship between the various temperatures along these minima schematically, as discussed in the text. We could not constrain the location of \Ttri and \Tc is our simulations at zero pressure due to the small temperature region between them and \Tt. \Ths is presented by the dashed red segment, the first-order transition region between \Tt and \Ttri is the solid black segment, \Ts is the dashed blue segment between \Ttri and \Tc, and \Tstar is the dotted violet high T segment.}
\end{figure}

Our results agree well with recent experiments: Bai et al. found that in BaTiO$_3$ films the maximum ECE occurs above \Tc and increases with applied field \cite{bai2011,bai2012}. Relaxors and relaxor ferroelectrics are expected to show similar behavior to what we see in \LNO (e.g.~Fig. 3 in Ref.~\onlinecite{kutnjak2007}). 

\begin{figure}[tbp]
\centering
\includegraphics[width=3.5in]{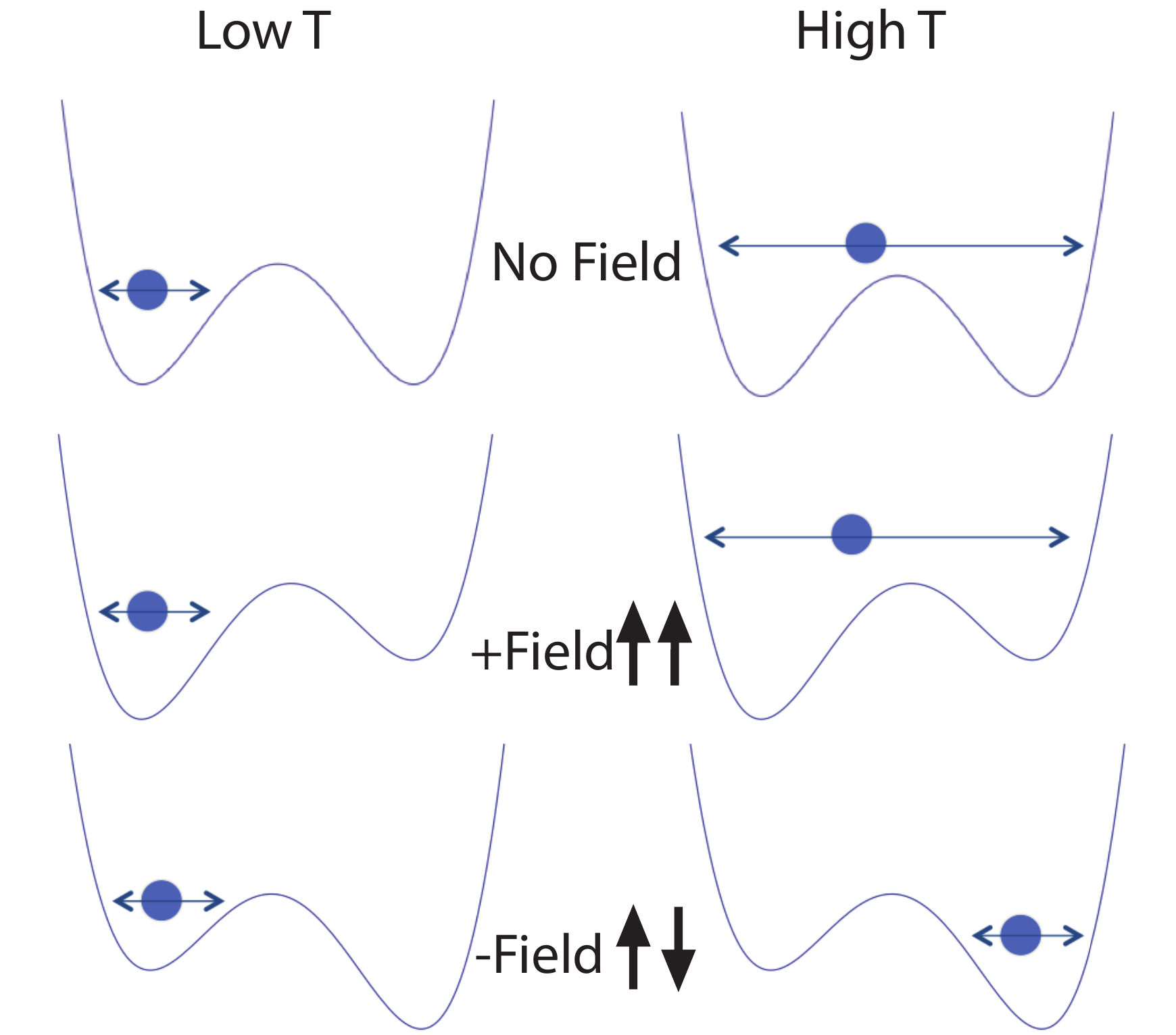}
\caption{\label{doublewells}The connection between the transition temperatures and switching below \Tt with an antiparallel applied field is shown in a simple schematic. These do not show the Landau free energy, but rather the potential energy versus local polarization. In all cases an activation barrier exists for changing the local polarization.}
\end{figure}

One might find it surprising at first that an apparently kinetic phenomenon, ferroelectric switching, is related to the thermodynamics at \Tt, and moreover, to the behavior high above the zero field \Tc where there is no phase transition at all.  We can understand this behavior from an atomistic perspective by considering a simple classical double well representing the underlying potential surface (Fig.~\ref{doublewells}).  The key is that the same double well potential surface underlies ferroelectric switching, \Tt, and the maximum in $\epsilon \propto C$:  In zero applied field, \Tt is given by coupled local double well potentials. In the displacive limit \Tt is the temperature where there is enough kinetic energy to traverse the hump in the underlying effective double wells. In an applied field the underlying potentials are asymmetric, and it takes a higher $T$ to go over the hump. There is no phase transition above the critical field, but the underlying potential surface is still asymmetric due to the field, and the susceptibility, which diverges at a true phase transition, instead peaks at the temperature where local modes start to hop over the hump. This is the Widom line. Below \Tt the effective potentials are asymmetric due to the coupling between local modes; the coupling can be thought of as a local Weiss field from the rest of the crystal.  When a field greater than the coercive field is applied, the polarization switches. In terms of the double well picture in Fig.~\ref{doublewells}, the field makes a population inversion, where the high energy state is occupied. At low temperatures or below the coercive field for a given temperature the atoms will be stuck in the high energy state; as temperature is raised  there will be a point at which the local modes all cross the hump and go into the lower energy state. So we see the origin of the connection between switching, \Tt, and the Widom line. In the order-disorder limit, with a huge hump that cannot be broached, the dynamics are governed by nucleation and growth, but we still expect to see a relationship between switching, \Tt, and the Widom line since the field dependence is still governed by the $E \cdot P$ term in the free energy.

This behavior for applied fields carries over as well to magnetism, and surprisingly does not seem to have been remarked upon in the literature. In experiments for magnetization of pure iron under applied magnetic fields, the Widom line is clearly seen about \Tc, and the Widom line comes in with a finite slope at \Tc \cite{liu2008}. This suggests that magnetic switching should also follow this line below \Tc, otherwise the Widom line would come in with zero slope. So homogeneous switching, \Tc, and the Widom line in ferromagnets is related in the same way to applied magnetic fields, as ferroelectrics behave in applied electric fields. 

Some have attributed the ECE to polarization disordering giving a resulting change in entropy \cite{Ponomareva2012}, but this simple view can be misleading. For example, an isolated supercooled liquid will crystallize with a concomitant increase in entropy, as required by the second law of thermodynamics, even though a crystal is more ordered than a liquid. Pressure decreases \Tc, and thus increases the ECE, not because it directly causes greater disorder (an entropic effect), but simply because the paraelectric phase is denser than the ferroelectric phase, and thus favored by pressure. The Maxwell relation, Eq.~\ref{maxwell1}, says everything necessary about the relationship of the ECE to the pyroelectric effect. Furthermore, this is an exact expression, and thus the ``direct" and ``indirect" approaches to the ECE must agree within the error of experiments or simulations, in contrast to Ref.~\onlinecite{Ponomareva2012}. It is also important to realize that the $\Delta S_{tr}$, the entropy change for the ferroelectric to paraelectric first-order phase transition is not the pertinent quantity for the ECE, but rather $(\partial S/\partial E)|_T$. Large $\Delta S_{tr}$ can give rise to a large ECE near \Tt, but only in the first-order transition region. We find larger ECE at higher fields and temperatures beyond \Ttri.

Our MD simulations with first-principles potentials on \LNO  give some universal conclusions about the electrocaloric effect (ECE). We expect the ECE to be maximal above \Tt, suggesting materials with \Tt below room temperature (or the operating temperature) would be optimal. We find a relationship between \Tt, the Widom line, and homogeneous switching, and suggest this may be universal among ferroelectrics, relaxors, multiferroics and ferromagnets. The ECE should be large in any insulator with a large, temperature dependent, dielectric susceptibility, not just ferroelectrics. We have shown the importance of the Widom line and the nature of the ECE above \Tt, the relationship between homogeneous switching and behavior above \Tt in an applied electric field, the continuity of the Widom line with the switching behavior below \Tt, the realization that ECE does not require a ferroelectric; the relationship of switching and the Widom line in ferromagnets as well as ferroelectrics, clarification of conditions to maximize the ECE in materials, the behavior of the ECE under pressure, and finally the application of an accurate  first-principles based shell-model to behavior in an applied electric field.

\begin{acknowledgments}
This work was supported as part of the EFree, an Energy Frontier Research Center funded by the U.S. Department of Energy, Office of Science, Office of Basic Energy Sciences under Award Number DE-SC0001057, and the Carnegie Institution of Washington. We thank P. Ganesh for indicating that the ECE $\ne 0$ at finite fields.We thank I. Naumov, J.F. Scott, Q. Peng, and M. Ahart for useful discussions. 
\end{acknowledgments}
\bibliography{rose}

\end{document}